\UseRawInputEncoding
\documentclass[journal]{IEEEtran}
\usepackage{amsfonts}
\usepackage{amssymb}
\usepackage{amsthm}
\usepackage{amsmath}
\usepackage{mathrsfs}
\usepackage{subfigure}
\usepackage{cite}
\usepackage{graphicx}
\usepackage{epstopdf}
\usepackage{enumerate}
\usepackage{color}
\usepackage{multirow}
\usepackage{booktabs}
\usepackage{algorithm}
\usepackage{algorithmic}

\renewcommand{\algorithmicrequire}{\textbf{Initialization:}}

\makeatletter

\newcommand\old[1]{}
\newtheorem{definition}{Definition}
\newtheorem{lemma}{Lemma}
\newtheorem{corollary}{Corollary}

\newtheorem{theorem}{Theorem}
\newtheorem{remark}{Remark}
\newtheorem{example}{Example}
\newtheorem{assumption}{Assumption}

\makeatother

\begin{document}

\date{\today}

\title{Design and Implementation of Data-driven Predictive Cloud Control System}
\author{Runze Gao, Yuanqing Xia$^{*}$, Li Dai, Zhongqi Sun
\thanks{R. Gao, Y. Xia, L. Dai and Z. Sun are with the School of Automation, Beijing Institute of Technology, Beijing 100081, China. ({\footnotesize {\em Corresponding
author: Yuanqing Xia}). Email address:
runze$\_$gao@bit.edu.cn (R. Gao), xia$\_$yuanqing@bit.edu.cn (Y. Xia), li.dai@bit.edu.cn (L. Dai), zhongqisun@bit.edu.cn (Z. Sun)
}}}
\markboth{manuscript for review}{}
\maketitle

\begin{abstract}
Nowadays, the rapid increases of the scale and complexity of the controlled plants bring new challenges such as computing power and storage for conventional control systems. Cloud computing is concerned as a powerful solution to handle the complex large-scale control missions using sufficient computing resources. However, the developed computing ability enables more complex devices and mass data being involved and thus the applications of model-based algorithms are constrained. Motivated by the above, we propose an original data-driven predictive cloud control system. To achieve the proposed system, a practical data-driven predictive cloud control platform rather than only a numerical simulator is established and together a cloud-edge communication scheme is developed. Finally, the verification of simulations and experiments as well as discussions demonstrate the effectiveness of the proposed system.
\end{abstract}

\begin{IEEEkeywords}
Cloud Control System, Cloud Computing, Data-driven Predictive Control
\end{IEEEkeywords}

\section{Introduction}\label{Introduction}
\begin{figure}[!htb]
  \centering
  \includegraphics[width=2.75in]{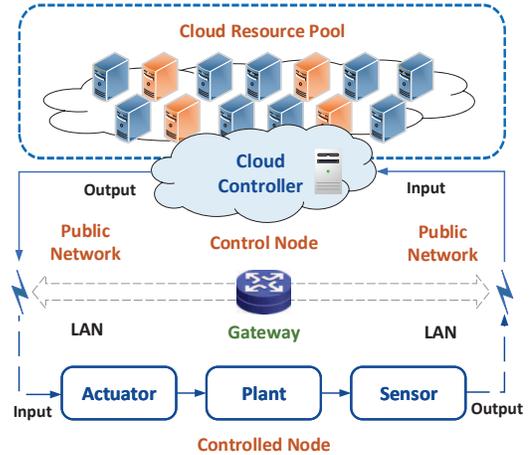}
  \caption{Structure of the Cloud Control System Model}
  \label{Structure of Cloud Control System Model}
\end{figure}

\IEEEPARstart{W}{ith} the development of Internet of Things and cyber-physical system, the scale and complexity of the controlled plants increase rapidly \cite{xia2012networked}. This brings challenges for control systems of which the computing resource is limited. Cloud computing is concerned as a powerful solution which outsources the missions to virtual remote servers to reduce the computation overhead \cite{marinescu2017cloud, sultangazin2020symmetries, alexandru2020cloud}. Cloud computing has advantages on sufficient computational resources, utilizing global information, handling large-scale data \cite{tanaka2017directed}, which has been used in multi-agents, smart grids and autonomous vehicles \cite{liu2017predictive, bera2014cloud, wang2021cloud}. Based on cloud computing, cloud control system (CCS) is a new paradigm to design and implement the control framework and services \cite{wang2021cloud, xia2015cloud, mahmoud2017interaction}. In CCS, a shared cloud resource pool consisting of CPU, memory, storages and other resources is used to replace the single controller in conventional control system as shown in Fig. \ref{Structure of Cloud Control System Model} \cite{xia2020cloud}. However, the increasing computing power brought by cloud computing enables more complex devices and large-scale data being involved in control systems. And it is hard even impossible to obtain the accurate physical model of every device in CCS \cite{xia2015cloud}. Meanwhile, the large-scale data are collected and stored but can not be fully utilized by model-based algorithms \cite{yin2015data}. Therefore, a cloud-based data-driven control system is indeed required to deal with control missions not relying on pre-established physical models.

Different from model-based algorithms, data-driven control serves as a model-free method, by skipping the modeling procedure and obtaining control variables directly from the input and output data \cite{qiu2017data}. However, data-driven control method requires larger scale data and brings greater computation, which cause time delays. Beside in this process, time delays also occur in the upload and download processes between the cloud controller and the controlled plants, which may affect the control quality especially with real-time requirement. Data-driven predictive control method provides a solution as generating a predictive control sequence and this sequence cloud be used to compensate the time delays \cite{huang2008dynamic}, which has been applied in networked control system \cite{xia2013data}. Thus, a data-driven control method is adopted here to reduce the affect of time delays in an cloud-based system and the data-driven predictive cloud control system is proposed.

Moreover, to verify the performance of the proposed system, a practical data-driven predictive cloud control platform rather than only a numerical simulator is required. To the best of the authors' knowledge, the desired platform does not yet exist. To establish the platform, some practical problems are raised. First, the design of the practical platform is needed. Second, the communication scheme between the cloud and edge devices occupies the significant role. For instance, the cloud controller and the controlled devices are in different network environments and the cloud controller has no knowledge about the hosts of the devices. How to build the cloud-edge communication bind at the beginning? What is more, the choice of communication protocol affects the control quality of the proposed system through time delays and packet losses rate. But the operations to reduce packet losses rate such as multiple authentication and data retransmission rise time delays \cite{kumar2012survey}. So what is the propensity of demand of the proposed system and how to choose the protocol based on the demand? The above problems are required to be solved. In this work, we focus on the design of a cloud-based framework of data-driven predictive control and the implementation of the practical platform. The main contributions are summarized as follows.

\begin{enumerate}
  \item An original data-driven predictive cloud control system is proposed to achieve desired control performance only relying on the input and output data via cloud computing and to compensate the time delays in cloud environment.
  \item A practical data-driven predictive cloud control platform rather than a numerical simulator is established, together with a cloud-edge communication scheme.
  \item The verification of simulations and experiments are preformed as well as the discussions are provided, which demonstrate the effectiveness of the proposed system.
\end{enumerate}

The remainder of this article is organized as follows. Section~\ref{Preliminary} introduces the preliminary of data-driven predictive control. Section~\ref{Data-driven Predictive cloud control system section} provides the architecture and algorithm of the data-driven predictive cloud control system. The practical platform is designed and established in Section~\ref{Cloud Control Experiment Platform Design and Verification}. The verification and discussions are provided in Section~\ref{Simulations and Experiments}. Finally, the conclusion and future work are presented.


\section{Preliminary}\label{Preliminary}

In this preliminary, a data-driven predictive control algorithm proposed in \cite{xia2013data} is recalled. Assume the collections of the inputs and outputs $\{ u(n), y(n), n = k-N, k-N+1,\ldots,k+N-1\}$ are available, where $k$ is the sampling time and $N \in \mathbb{N}^{+}$. The column vectors over a time horizon consisting of the inputs and outputs are defined as
\begin{equation}\label{e1}
  \small{u_f(k) \!=\! \left[\!
          \begin{array}{c}
            u(k) \\
            u(k\!+\!1) \\
            \vdots \\
            u(k\!+\!N\!-\!1) \\
          \end{array}\!
        \right]\!, \, y_f(k) \!=\! \left[\!
          \begin{array}{c}
            y(k) \\
            y(k\!+\!1) \\
            \vdots \\
            y(k\!+\!N\!-\!1) \\
          \end{array}\!
        \right]\!, \, }
\end{equation}
\begin{equation}\label{e2}
  \small{u_{p}(k) \!=\! \left[\!
          \begin{array}{c}
            u(k-\!N) \\
            u(k-\!N\!+\!1) \\
            \vdots \\
            u(k\!-\!1) \\
          \end{array}\!
        \right]\!, \, y_{p}(k) \!=\! \left[\!
          \begin{array}{c}
            y(k\!-\!N) \\
            y(k\!-\!N\!+\!1) \\
            \vdots \\
            y(k\!-\!1) \\
          \end{array}\!
        \right]\! \,}
\end{equation}
where the subscript $``\emph{p}"$ stands for $``$past$"$ and $``\emph{f}"$ means $``$future$"$. Further, let
\begin{equation}\label{e3}
  \small{w_p(k) = \left[
             \begin{array}{c}
               y_{p}(k) \\
               u_{p}(k) \\
             \end{array}
           \right]\!.}
\end{equation}

Using regression analysis approach \cite{huang2008dynamic}, we obtain
\begin{equation}\label{e4}
  \small{y_{f}(k)=L_{w}w_{p}(k)+L_{u}u_{f}(k)+L_{e}e_{f}(k)}
\end{equation}
where $e_{f}(k)$ is the future column vector, similar with $u_f(k)$ and $y_f(k)$, consisting of the white noise $\{ e(n), n = k, k+1,\ldots,k+N-1\}$ and $L_w$, $L_u$, $L_e$ are the corresponding coefficient matrices with proper dimensions. Since $e_f(k)$ is a white noise vector, the predicted future output sequence, $\widehat{y}_f(k)$ is given as
\begin{equation}\label{e5}
  \small{\widehat{y}_{f}(k)=L_{w}w_{p}(k)+L_{u}u_{f}(k).}
\end{equation}

To calculate $\widehat{y}_{f}(k)$, $L_{w}$ and $L_{u}$ are required. In this data-driven algorithm, Hankel matrices are applied to obtain $L_{w}$ and $L_{u}$. Suppose the collections of the input and output data $\{ u(n), y(n), n = 1,2,\ldots,2N+j-1\}$ are available, where $j \in \mathbb{N}^{+}$ . The Hankel matrices of the inputs, represented as $U_p$ and $U_f$, with $N$-block rows and $j$-block columns are defined as
\begin{eqnarray}\label{e6}
  \small{\!\!\!\!\!\!U_p \!\!\!\!}&=& \small{\!\!\!\!\!\!\left[
            \begin{array}{cccc}
              \!\!u(0) & \!\!u(1) & \!\!\ldots & \!\!u(j-1) \\
              \!\!u(1) & \!\!u(2) & \!\!\ldots & \!\!u(j) \\
              \!\!\vdots & \!\!\vdots & \!\!\ddots & \!\!\vdots \\
              \!\!u(N-1) & \!\!u(N) & \!\!\ldots & \!\!u(N+j-2) \\
            \end{array}
          \right]}, \\
  \small{\!\!U_f \!\!\!\!}&=& \small{\!\!\!\!\!\!\left[
            \begin{array}{cccc}
              \!\!\!u(N) & \!\!\!u(N+1) & \!\!\!\ldots & \!\!\!u(N+j-1) \\
              \!\!\!u(N+1) & \!\!\!u(N+2) & \!\!\!\ldots & \!\!\!u(N+j) \\
              \!\!\!\vdots & \!\!\!\vdots & \!\!\!\ddots & \!\!\!\vdots \\
              \!\!\!u(2N-1) & \!\!\!u(2N) & \!\!\!\ldots & \!\!\!u(2N+j-2) \\
            \end{array}
          \right]}
\end{eqnarray}
The Hankel matrices of the outputs are written likewise, denoted as $Y_p$ and $Y_f$. Further, let
\begin{equation}\label{e8}
  \small{W_p=\left[
           \begin{array}{c}
             Y_p \\
             U_p \\
           \end{array}
         \right].}
\end{equation}

Then we obtain
\begin{eqnarray}
  \small{U_f \!\!\!}&=&\small{\!\!\! \left[\!\!
          \begin{array}{cccc}
            u_f(N)\!\! & u_f(N+1)\!\! &\! \ldots & \!\!u_f(N+j-1) \\
          \end{array}\!\!
        \right]} \label{e9}\\
  \small{Y_f \!\!\!}&=&\small{\!\!\! \left[\!\!
          \begin{array}{cccc}
            y_f(N)\!\! & y_f(N+1)\!\! & \ldots &\!\! y_f(N+j-1) \\
          \end{array}\!\!
        \right]} \label{e10}\\
 \small{ W_p \!\!\!}&=&\small{\!\!\! \left[\!\!
          \begin{array}{cccc}
            w_p(N)\!\! \!& w_p(N+1)\!\! \!& \ldots \!&\!\! w_p(N+j-1) \\
          \end{array}\!\!
        \right]} \label{e11}
\end{eqnarray}

Using (\ref{e9})-(\ref{e11}) to extend (\ref{e4}), we obtain the Hankel matrix of the future output data with the same coefficient matrices
\begin{equation}\label{e12}
  \small{Y_{f}=L_{w}W_{p}+L_{u}U_{f}+L_{e}E_{f}}
\end{equation}
where $E_f = [e_f(N) \ e_f(N) \ \ldots \ e_f(N+j-1)]$ and the white noise $\{ e(n), n = 1, 2,\ldots,2N+j-1\}$ are available. Similarly with (\ref{e5}), the Hankel matrix of the predicted output, $\widehat{Y}_{f}$ is computed as
\begin{equation}\label{e13}
  \small{\widehat{Y}_{f}=L_{w}W_{p}+L_{u}U_{f}.}
\end{equation}

By solving the following least squares problem
\begin{equation}\label{e14}
  \small{\min_{L_{w},L_u}\|Y_{f}-\left[
                           \begin{array}{cc}
                             \!\!\!L_{w} & \!\!\!L_{u} \\
                           \end{array}\!\!
                         \right]\!\left[
                                  \!\begin{array}{c}
                                    W_{p} \\
                                    U_{f} \\
                                  \end{array}\!\!
                                \right]\!
\|^{2}_{F}}
\end{equation}
where $\|\cdot\|_{F}$ means $F$-norm, the coefficient matrices $L_w$ and $L_u$ can be computed by
\begin{eqnarray}\label{e15}
  \small{\left[
                           \!\!\begin{array}{cc}
                            L_{w} & \!\!\!\!L_{u} \\
                          \end{array}\!\!
                        \right] \!\!\!\!}&=& \small{\!\!\!\!Y_{f}\left[
                                 \!\!\begin{array}{c}
                                   W_{p} \\
                                   U_{f} \\
                                 \end{array}\!\!
                               \right]^{\dag}} \nonumber\\
   \!\!\!\!&=& \small{\!\!\!\!Y_{f}\left[
                                                       \!\begin{array}{cc}
                                                         W_{p}^{T} & \!\!\!U_{f}^{T} \\
                                                       \end{array}\!\!
                                                     \right]\!\left[
                                                \begin{array}{cc}

                                                  \!\!\!\left[
                                                   \begin{array}{c}
                                                     \!\!W_{p} \\
                                                    \!\!U_{f} \\
                                                  \end{array}\!\!\!
                                                  \right]
                                                  & \!\!\!\!\!\!\left[
                                                       \begin{array}{cc}
                                                         \!W_{p} & \!\!\!U_{f} \\
                                                       \end{array}\!\!
                                                     \right]
                                                    \\

                                                \end{array}\!\!\!
                                              \right]^{-1}}.
\end{eqnarray}
\begin{figure*}[!htbp]
  \centering
  \includegraphics[width=5.5in, height=3.95in]{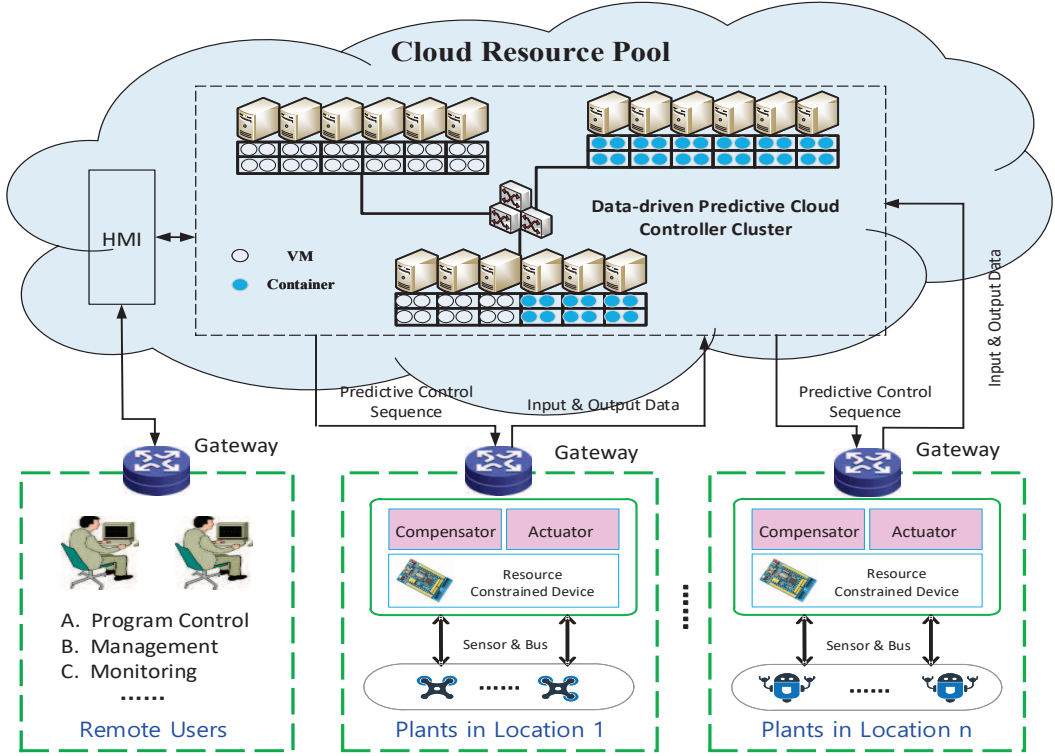}\\
  \caption{Architecture of the Data-driven Predictive Cloud Control System}\label{Architecture of Data-driven Predictive Cloud Control System}
\end{figure*}

Since $\small{\left[
                                 \!\!\begin{array}{c}
                                   W_{p} \\
                                   U_{f} \\
                                 \end{array}\!\!
                               \right]}$ is not a square matrix, the Morre-Penrose pesudo-inverse $\dag$ \cite{zhang2017matrix} is used here.

Finally, consider an unconstrained MPC problem with the objective function
\begin{equation}\label{e16}
  \small{J=(r_{f}-\widehat{y}_{f}(k))^{T}(r_{f}-\widehat{y}_{f}(k))+u_{f}(k)^{T}(\lambda I)u_{f}(k)}
\end{equation}
where $r_f$ is the reference value of the output, $\lambda$ is an adjustable coefficient and $I$ is an unit matrix. Substituting (\ref{e5}) into (\ref{e16}) and taking derivative with respect to $u_f$, the optimal data-driven predictive control law without constraints can be computed as
\begin{equation}\label{e17}
  \small{u_{f}(k)=(\lambda I+L_{u}^{T}L_{u})^{-1}L_{u}^{T}(r_{f}-L_{w}w_{p}(k)).}
\end{equation}

\section{Data-driven Predictive Cloud Control system}\label{Data-driven Predictive cloud control system section}

The proposed data-driven predictive cloud control system is designed in this section. First, the architecture of this proposed system is established. Then, to reduce the influence of cloud-edge time delays, a cloud-edge compensator is designed. Finally, an algorithm corresponding to the data-driven predictive cloud control system is proposed.

\subsection{System architecture}

Fig. \ref{Architecture of Data-driven Predictive Cloud Control System} sketches the architecture of the proposed system. This architecture consists of three layers, which are cloud platform layer, edge layer and user layer. Cloud platform layer is on the upper side of this architecture. This layer takes charge of collecting useable resource and creating computing node such as virtual machine (VM) and container \cite{pahl2017cloud}, which would be used as the data-driven cloud controller for the edge plant. The input and output data generated by the edge plant are transmitted to the cloud controller and then processed by the imported data-driven predictive control algorithm. Finally, the optimal predictive control sequence are obtained and sent to the edge node. In addition, the cloud controller is connected to the remote user via human-machine interface (HMI).

Edge layer is on the below right of this architecture. There are multiple edge nodes deployed in different locations, where the output data are generated by the controlled plant and collected by the sensors. The output data are packed with the actuated input data at the same control period and the obtained packet is sent to the cloud controller together via gateway. In the next period, the terminal node receives the computed optimal predictive control sequence. Since the data communication are based on network and the data-driven algorithm also costs time, time delays between the cloud controller and edge plant occur in the proposed system. To reduce the influence of time delays, a cloud-edge compensator is designed to pick a proper control variable from the sequence which will be provided for the actuator. User layer is on the below left of this architecture. In this layer, the interactive data are applied in different functions such as the program control, management and monitoring.

\subsection{Cloud-edge compensator design}
\begin{figure}[!htb]
  \centering
  \includegraphics[width=3.3in, height=2.6in]{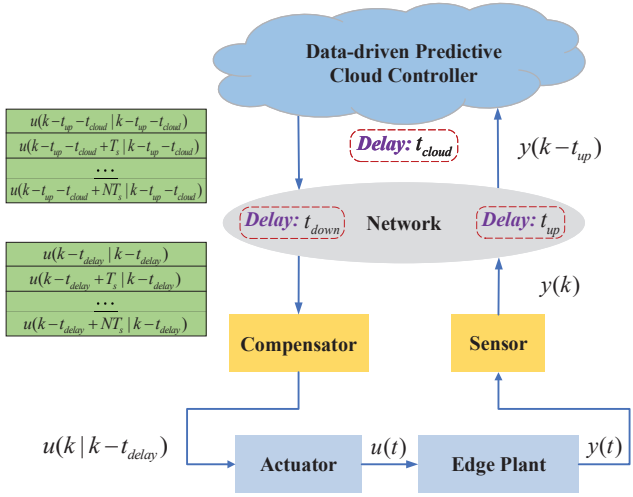}
  \caption{Time Delays Between the Cloud Controller and Edge Plant}
  \label{Time Delay Between Cloud Controller and Edge Plant}
\end{figure}

Since network and cloud computing node are involved, time delays become a serious problem to be solved. The total time delay $t_{delay}$ is defined as
\begin{equation}
  t_{delay} = t_{up} + t_{cloud} + t_{down}
\end{equation}
where $t_{up}$, $t_{cloud}$ and $t_{down}$ represent the upload, cloud computing and download delays, respectively. The detailed processes are described in Fig. \ref{Time Delay Between Cloud Controller and Edge Plant}.
First, the sensors collect the output data $y(k)$ at the discrete sampling time $k$. Then the data are uploaded to the cloud controller via network, of which the cost time is $t_{up}$. When the data arrives at the cloud controller, it becomes $y(k-t_{up})$. Similarly, by providing the sampling period as $T_s$, the predictive control sequence is written as $\{ u(k-t_{up}-t_{cloud}+iT_s|k-t_{up}-t_{cloud}), i = 0,1,\ldots, N \}$. Then, when this sequence is downloaded to the edge node, it becomes $\{ u(k-t_{delay}+iT_s|k-t_{delay}), i = 0,1,\ldots, N \}$.

To reduce the influence of time delays, a cloud-edge compensator is designed and deployed. When the output data leaves the edge node, the time is recorded as $t_1$. Similarly, the time of the control sequence arriving at edge node is recorded as $t_2$. Thus the total time delay is computed as $t_{delay} = t_2 - t_1$ and the standard delay unit can be obtain as $\tau = [\frac{t_{delay}}{T_s}]$ where the square bracket $[\,\, \cdot \,\, ]$ indicates if the number is an integer, it would be kept and otherwise it would be rounded. Finally, the compensator picks $u(k-t_{delay}+\tau T_s|k-t_{delay})$, the $\tau$-th variable in the received sequence as the control input to reduce the influence of the delays, since this variable is approximately equal to $u(k|k-t_{delay})$. In addition, the influence of packet losses can be also compensated. When the edge client receives no signal, the actuator would use the predictive control sequence of the last period. After the edge client receives data again, the compensator would return to use the present sequence.

\subsection{Algorithm design}
\begin{figure}[!htb]
  \centering
  \includegraphics[width=3.25in, height=2.6in]{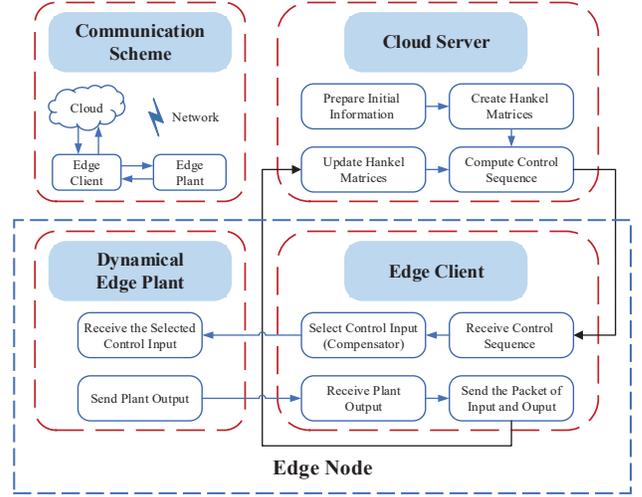}
  \caption{Data Flow of the Data-driven Predictive Cloud Control Algorithm}
  \label{Data flow of the Data-driven Predictive Cloud Control Algorithm}
\end{figure}

In this subsection, the data-driven predictive cloud control algorithm is designed (see Algorithm \ref{Data-driven Predictive cloud control system Algorithm}). As shown in Fig. \ref{Data flow of the Data-driven Predictive Cloud Control Algorithm}, the data flow of this algorithm is conducted in three subparts, which are cloud server, edge client and edge plant. In this data flow, a communication scheme among the three parts is required to support the data transmit, which is established in subsection \ref{Cloud communication system design and verification}. The first part of this data flow is in the cloud server which takes charges of the computation of the data-driven algorithm. First, this cloud server prepares a certain amount of initial input and output data in advance by applying other control method such as PID to the controlled plant. Then the initial data are stored in Hankel matrices such as $U_f$, $U_p$ and etc. The generated Hankel matrices are used to compute the optimal predictive control sequence, which is be sent to the edge node.

\begin{algorithm*}[!t]
    \renewcommand{\algorithmicrequire}{\textbf{Input:}}
	\renewcommand{\algorithmicensure}{\textbf{Output:}}
    \caption{\textbf{Data-driven Predictive Cloud Control System Algorithm}}
    \label{Data-driven Predictive cloud control system Algorithm}
    \textbf{Input:}
    The scale parameters $N$ and $j$ of Hankel matrices, sampling interval $T_s$ and reference value $r_f$.\\
    \textbf{Output:} Optimal data-driven predictive signal $u(k)$ provided by the cloud controller.
    \begin{algorithmic}[1]
        \STATE \textbf{\textit{Preparing Information}}
        \begin{enumerate}
          \item \textit{Set parameters:} First, set the scale parameters $N$ and $j$ to determine the size of Hankel matrices. The sample interval $T_s$ and reference $r_f$ are also need be set.
          \item \textit{Collect data:} Generate a series of initial control inputs
           $\{u(0),u(1),\ldots,u(2N+j-1)\}$ by a certain algorithm such as PID. Then measure the positions of the plant $\{y(0),y(1),\ldots,y(2N+j-1)\}$ as outputs by sensors.
          \item \textit{Create Hankel matrices:} Create the initial Hankel matrices of the collected input and output data.
        \end{enumerate}
        \STATE \textbf{\textit{Cloud Server}}
        \begin{enumerate}
          \item \textit{Intermediate variables:} At sampling time $k$, calculate the middle variables $L_u$ and $L_\omega$, $\left[
                           \!\!\begin{array}{cc}
                            L_{w} & \!\!\!\!L_{u} \\
                          \end{array}\!\!
                        \right] = Y_{f}\left[
                                                       \!\begin{array}{cc}
                                                         W_{p}^{T} & \!\!\!U_{f}^{T} \\
                                                       \end{array}\!\!
                                                     \right]\!\left[
                                                \begin{array}{cc}

                                                  \!\!\!\left[
                                                   \begin{array}{c}
                                                     \!\!W_{p} \\
                                                    \!\!U_{f} \\
                                                  \end{array}\!\!\!
                                                  \right]
                                                  & \!\!\!\!\!\!\left[
                                                       \begin{array}{cc}
                                                         \!W_{p} & \!\!\!U_{f} \\
                                                       \end{array}\!\!
                                                     \right]
                                                    \\

                                                \end{array}\!\!\!
                                              \right]^{-1}\!\!\!$ which is mentioned in section \ref{Preliminary} from the newest Hankel matrices;
          \item \textit{Control sequence:} Obtain the control sequence $u_f(k) = [u(k|k) \;\; u(k+1|k) \;\; \ldots \;\; u(k+N|k)]$ as the formula (17) by using $L_u$ and $L_\omega$;
          \item \textit{Package and send:} Pack the control sequence $u_f$ and present time $t_{1}$ recorded by a standard clock. Then send the information to the edge client together.
        \end{enumerate}
        \STATE \textbf{\textit{Edge Client}}
        \begin{enumerate}
          \item \textit{Receive data and calculate delay unit:} The edge client receives the control sequence $u_f$ and records the standard time $t_{2}$ which is equal to $t_{1}+t_{delay}$. Compute the delay unit $\tau = [\frac{t_{delay}}{T_s}]$;
          \item \textit{Compensate delay and connect actuator:} Choose the $\tau$-th value in the received control sequence $u_f(k)$ as the real input $u(k)$. Apply $u(k)$ to the dynamic system and detect the corresponding output $y(k)$;
          \item \textit{Package and send:} Pack the chosen input $u(k)$ and the corresponding $y(k)$ as a combined packet. Then send the packet to cloud server for the optimization at next step.
        \end{enumerate}
        \STATE \textbf{\textit{Update and Repeat}}
        \begin{enumerate}
          \item \textit{Update Hankel matrices:} The cloud server receives the latest packet of $u(k)$ and $y(k)$, and updates the Hankel matrices by removing the earliest data as well as adding the newest data to keep the fixed size $N$;
          \item \textit{Repeat:} Repeat the course of \textbf{\textit{Cloud Server}} and continue the circulate until receiving a stop command.
        \end{enumerate}
    \end{algorithmic}
    \label{algorithm}
\end{algorithm*}

The data flow in the edge node is made up by the edge client and dynamical plant subparts. Since the computing resource in edge node is limited, this edge client is created only to deliver data between the cloud server and edge plant based on the limited resource. This edge client keeps listening for the predictive control sequence from the cloud server and computes the cloud-edge delay. A proper control variable is selected from this sequence by the compensator, which would be sent to the dynamical edge plant. After applying this control variable to the edge plant, the corresponding output data are collected by the sensors and then sent to the edge client. Finally, the input and output data are packed together and then uploaded to the cloud server to update the Hankel matrices and compute a new control sequence.

The algorithm is designed by four steps: (1) prepare initial information; (2) cloud server: data process and control sequence computation; (3) edge client: compensate time delays and execute control input; (4) data update with feedback. The four steps are performed in the system as a circulation. In this system, we implicitly identify the model of dynamical system in real time, yet the model is not established throughout the course of control. At each sampling time, the model is updated by real-time data. Then, the MPC method is employed in the implicit dynamical model. In fact, this model is cancelled as an intermediate process and the control process only relies on data. Therefore, it is named as data-driven predictive cloud control system.

\begin{remark}
Clock synchronization is also a considerable concern between the cloud server and edge client. In this work, we assume all the nodes are set as the national standard time.
\end{remark}

\begin{figure}[!htb]
  \centering
  \includegraphics[width=3.35in, height=1.82in]{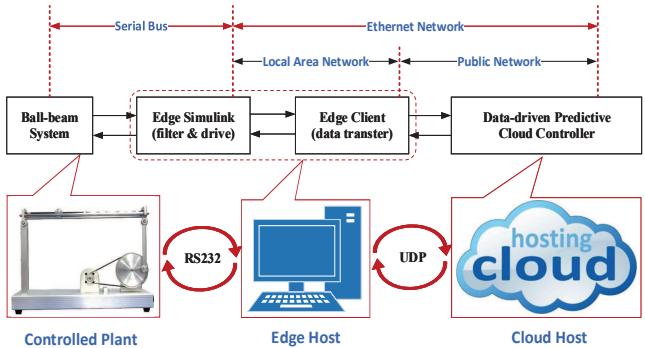}
  \caption{Structure of the Data-driven Predictive Cloud Control Platform}
  \label{Structure of the Data-driven Predictive Cloud Control Platform}
\end{figure}

\section{Cloud Control Experiment Platform Design}\label{Cloud Control Experiment Platform Design and Verification}

In this section, a data-driven predictive cloud control platform is established. First, the design of this cloud control experiment platform is provided. Then, a cloud-edge communication scheme is designed and verified, in which the problems of IP transfer and protocol choice are solved.

\begin{figure}[!htb]
  \centering
  \includegraphics[width=3in, height=1.82in]{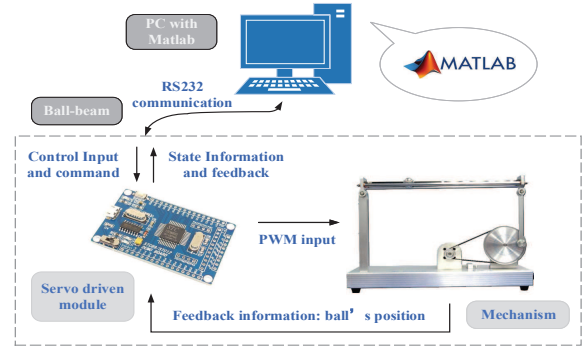}
  \caption{Structure and Data Flow of the Edge Node}
  \label{Structure and Data Flow of the Edge Node}
\end{figure}

\subsection{Cloud control platform design }

Fig. \ref{Structure of the Data-driven Predictive Cloud Control Platform} depicts the structure of the data-driven predictive cloud control platform. For instance, the controlled plant is set as an ball-beam system which is described in section \ref{Ball-beam system}. In this structure, the controlled plant and edge client are deployed in the laboratory while the cloud controller works in public cloud. The edge client consists of a data transfer and a Matlab module, which is linked to the plant. In this client, the data transfer and Matlab module are connected directly. The client transmits data with the cloud across public network while delivering inside data via local area network (LAN).

The edge node consisting of the client and controlled plant is shown in Fig. \ref{Structure and Data Flow of the Edge Node}. This plant transmits the feedback data to a servo drive module. This drive module exchanges data with a edge computer with Matlab via RS232 interface. The exchanged data includes the state, feedback data and the control command. This client exchanges the edge information and control signals with the cloud controller. Finally, the drive module executes the control command to the plant by PWM signals.

\subsection{Cloud-edge communication system design}\label{Cloud communication system design and verification}

In this platform, the controller and plant are deployed in different network environments where the problems of IP transfer and protocol arise. Thus, a real-time, stable and high-quality communication system is required.

\noindent
\rule[0mm]{8.78cm}{0.2mm}

\emph{\textbf{Definition:}}

\textit{Netwok Tuple:} a tuple consisting of IP address and port, formed as \textit{(host:port)} or \textit{(host, port)}.

\textit{Internal Tuple:} network tuple of device in LAN.

\textit{Mapped Tuple:} network tuple mapped from an internal tuple by NAPT technology.

\textit{External Tuple:} network tuple of a device in public network.

\noindent
\rule[0mm]{8.78cm}{0.2mm}

\vspace{0.1cm}\noindent\emph{B1. IP transfer scheme design}\vspace{0.1cm}

\begin{figure}[htb]
  \centering
  \includegraphics[width=3.25in]{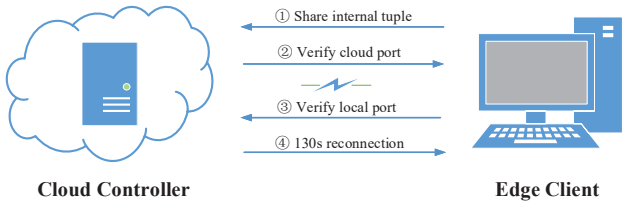}
  \caption{Cloud-Edge Communication System Bind Algorithm}
  \label{Cloud-Edge communication system bind algorithm fugure}
\end{figure}

Since the amount of public IP addresses is limited and each device running in edge environment also needs an IP address, network and port translation (NAPT) technology is adopted in LAN. In a edge node, all the devices share the same public IP and this public IP is mapped into different local IPs for the devices in LAN. However, the cloud controller has no knowledge about the hosts of the edge devices at the beginning. Thus, a special IP transfer scheme is designed as shown in Fig. \ref{Cloud-Edge communication system bind algorithm fugure} and is described in Algorithm \ref{Cloud communication system bind algorithm}.

In this scheme, the edge plant sends an information group consisting of the input and output signals and internal tuple to the cloud controller. Then, the cloud controller obtains the network host of this local plant. Furthermore, when the message is received by the cloud controller or edge plant, the receiver would check the attached port of the sender. If the ports of the receiver and sender are matched, the acknowledge process is completed. In addition, \cite{ma2017engineering} suggests that the communication bind can continue for 130s. If a period is more than 130s, the bind would be cancelled and built again.

\vspace{0.1cm}\noindent\emph{B2. Communication protocol choice}\vspace{0.1cm}

Besides the IP translation, protocol choice is another critical question. UDP and TCP are two most widely-used protocols in network communication \cite{kumar2012survey}. The main difference is that TCP can provide more reliable service for transmission, whereas UDP with high transmission speed can satisfy the requirement of real-time work. UDP is a message-oriented method with a simpler data flow control scheme which only requires IP address and port of receiver. Because of it’s simper scheme, UDP can deal with the transmission task with higher speed, smaller time delay. In brief, it is more efficient than TCP.

\begin{figure}[!htb]
  \centering
  \includegraphics[width=2.95in, height=0.85in]{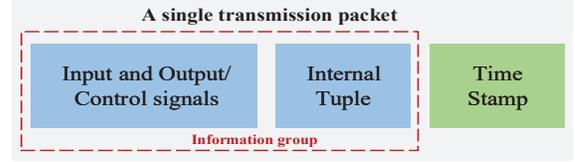}
  \caption{Structure of a Single Tramsmission Packet}
  \label{Structure of a Single Tramsmission Packet}
\end{figure}

\begin{algorithm}[!h]
    \renewcommand{\algorithmicrequire}{\textbf{Input:}}
	\renewcommand{\algorithmicensure}{\textbf{Output:}}
    \caption{\textbf{Cloud-Edge Communication System Bind Algorithm}}
    \label{Cloud communication system bind algorithm}
    \textbf{Problem:}
    Cloud controller cannot find the internal IP address of local device.\\
    \textbf{Goal:} Build the connection between the cloud and edge, and then transmit information.
    \begin{algorithmic}[1]
        \STATE \textbf{\textit{Connection establishment:}}
        \begin{enumerate}
          \item The edge device packages state information and internal tuple containing its host and port, and send the original information to the cloud controller.
          \item The cloud controller receives the package and learns the internal IP address of the edge device.
        \end{enumerate}

        \STATE \textbf{\textit{Identify verification:}}
         \begin{enumerate}
          \item In a one-way transmission, the cloud or edge side sends a message to another side.
          \item The receiver verifies whether the received port is matched with its own one.
        \end{enumerate}

        \STATE \textbf{\textit{Connection Maintenance:}}\\
        One bind maintenances for 130s. If the period is timeout, new bind would be built.
    \end{algorithmic}
    \label{algorithm}
\end{algorithm}

Time delays and packet losses are two most pivotal indexes of a communication system. Although TCP is more reliable, the transmission speed and single packet size are limited by its complex data flow manage scheme. As shown in the next subsection, UDP can deal with the time delays and packet losses well and thus it can offer a real-time, stable and high-quality transmission service.
\begin{table}[h]
    \centering
    \caption{Internet delay test(second)}
    \label{Internet delay test}
    \begin{tabular}{c|c|c|c|c}
    \hline
    \hline
    \textbf{\hspace{0.6cm} Time\hspace{0.6cm}}  &\textbf{\hspace{0.4cm} Region\hspace{0.4cm}}  &\textbf{Max}  &\textbf{Min}  &\textbf{Mean}\\
    \hline
    \multirow{2}{*}{8:00$\sim$10:00}  &\hspace{0.4cm}Beijing\hspace{0.4cm}  &0.0255  &0.0055  &0.0062\\
    \cline{2-5}
                                      &\hspace{0.4cm} Guangzhou\hspace{0.4cm}  &0.0235 &0.0185 &0.0218\\
    \hline
    \multirow{2}{*}{11:00$\sim$13:00}  &\hspace{0.4cm}Beijing\hspace{0.4cm}  &0.0235  &0.0075  &0.0149\\
    \cline{2-5}
                                       &\hspace{0.4cm} Guangzhou\hspace{0.4cm}  &0.0255  &0.0190  &0.0234\\
    \hline
    \multirow{2}{*}{14:00$\sim$16:00}  &\hspace{0.4cm}Beijing\hspace{0.4cm}  &0.0460  &0.0025  &0.0054\\
    \cline{2-5}
                                       &\hspace{0.4cm} Guangzhou\hspace{0.4cm}  &0.0375  &0.0185  &0.0209\\
    \hline
    \multirow{2}{*}{17:00$\sim$19:00}  &\hspace{0.4cm}Beijing\hspace{0.4cm}  &0.0620  &0.0025  &0.0211\\
    \cline{2-5}
                                       &\hspace{0.4cm} Guangzhou\hspace{0.4cm}  &0.0700  &0.0175  &0.0225\\
    \hline
    \hline
    \end{tabular}
    \end{table}

\subsection{Cloud-edge communication system verification}
To verify the performance of cloud-edge communication system, four groups of experiments are performed. With the cloud controllers deployed in Beijing and Guangzhou, respectively, communication delays are tested at different times in a day. The results are listed in Table \ref{Internet delay test}. The results show the average delay is nearly 0.02s and the maximum delay is less than $0.1s$. In the meanwhile, the range of packet losses rate is recorded from $0.2\%$ to $0.6\%$. The indexes of delay and packet losses rate can satisfy the requirement of the data-driven predictive cloud control system.
\begin{figure}[!htb]
  \centering
  \includegraphics[width=3in]{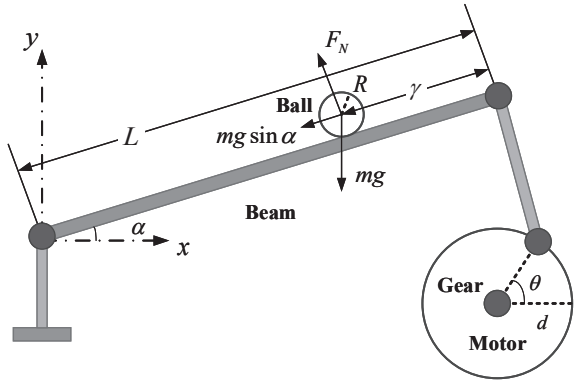}
  \caption{Model of the Ball-beam System}
  \label{Model of the Ball-beam System}
\end{figure}

\begin{table}[!htb]
    \centering
    \caption{Symbol description and parameter}
    \label{tab1}
    \begin{tabular}{c | ccc | c}
    \hline\hline
    \textbf{Symbol}        & &\textbf{Description} &     & \textbf{Parameter}\\
    \hline
    $L$        & &Length of the bean  &     & 0.4 $m$\\
    \hline
    $d$        & &Radius of the gear  &     & 0.04 $m$\\
    \hline
    $R$ 	 & &Radius of the ball &    & 0.015 $m$\\
    \hline
    $J_b$      &  &Moment of inertia of the ball &    & $9.9\times10^{-6}$ $kg\cdot m^{2}$\\
    \hline
    $m$          &   &Mass of the ball &    & 0.11 $kg$ \\
    \hline
    $g$	    & &Acceleration of gravity  &   & 9.81 $m/s^{2}$\\
    \hline
    $F_N$	    & &Support force on the ball  &   &$-$\\
    \hline
    \hline
    \end{tabular}
    \label{symbol description}
\end{table}
\section{Simulations and Experiments}\label{Simulations and Experiments}
In this section, a series of simulations and experiments are carried out based on an ball-beam system. First, the model of the ball-beam system is formulated. Second, we detect the feasibility of the cloud control platform by PID algorithm. Then the data-driven predictive cloud control algorithm (\textbf{Algorithm \ref{Data-driven Predictive cloud control system Algorithm}}) is applied to this platform. Two simulations are conducted to test the cloud-edge compensator. Finally, two simulations and two experiments are executed to verify the effectiveness of the data-driven predictive cloud control system.

\subsection{Ball-beam system}\label{Ball-beam system}
The structure of the ball-beam system is shown in Fig. \ref{Model of the Ball-beam System}. The main variables and parameters of this ball-beam system are listed in Table \ref{symbol description}. By choosing the ball position $\gamma$ and beam angle $\theta$ as the generalized position coordinates for this system \cite{rahmat2017application}, the Lagrangian equation of motion is given by:
\begin{equation}\label{e19}
  \small{0 = (\frac{J_b}{R^{2}}+m)\ddot{\gamma}+mg\sin\alpha-m\gamma\dot{\alpha}^{2}}
\end{equation}
where $\ddot{\gamma}$ is the acceleration of the ball, $\alpha$ is the beam angle and $\dot{\alpha}$ is the angular velocity of the beam angle. When this system approaches the stable point $\alpha=0$, the local linearization of (\ref{e19}) can be obtained. Since $\alpha$ is small at this point,  $\dot{\alpha}\approx0$ and $\sin\alpha=\alpha$. Therefore, the linear approximation of this system is given as
\begin{equation}\label{e20}
  \small{\ddot{\gamma} = -\frac{mg}{\frac{J_b}{R^{2}}+m}\alpha.}
\end{equation}

The beam angle $\alpha$ can be expressed related to the gear angle $\theta$ by approximating linear equation $\alpha L= \theta d$. Thus the relation between $\ddot{\gamma}$ and $\theta$ is
\begin{equation}\label{e21}
  \small{\ddot{\gamma} = -\frac{mdg}{L(\frac{J_b}{R^{2}}+m)}\theta.}
\end{equation}

In this system, the motor gear angle $\theta$ is set as the control input $u$. Letting the state be $x=[x_1\ x_2\ x_3\ x_4]^{T}=[\gamma\;\;\dot{\gamma}\;\;\theta\;\;\dot{\theta}]^{T}$ and output be $y = \gamma$, the state-space model is obtained as
\begin{eqnarray}
  \small{\left[\!
    \begin{array}{c}
      \dot{x}_1 \\
      \dot{x}_2 \\
      \dot{x}_3 \\
      \dot{x}_4 \\
    \end{array}\!
  \right] \!\!\!\!\!\!}&=&\small{\!\!\!\!\!\! \left[\!
            \begin{array}{cccc}
              0 & 1 & 0 & 0 \\
              0 & 0 & -\frac{mdg}{L(\frac{J_b}{R^{2}}+m)} & 0 \\
              0 & 0 & 0 & 1 \\
              0 & 0 & 0 & 0 \\
            \end{array}\!
          \right]\!\!\!\left[\!
    \begin{array}{c}
      x_1 \\
      x_2 \\
      x_3 \\
      x_4 \\
    \end{array}\!
  \right]\!\!+\!\!\left[\!
    \begin{array}{c}
      0 \\
      0 \\
      0 \\
      1 \\
    \end{array}\!
  \right]\!\!u} \\
  \small{y \!\!\!\!\!\!}&=&\small{\!\!\!\!\!\! \left[\!
      \begin{array}{cccc}
        1 & 0 & 0 & 0 \\
      \end{array}\!
    \right]\!\!\left[\!
    \begin{array}{c}
      x_1 \\
      x_2 \\
      x_3 \\
      x_4 \\
    \end{array}\!
  \right].}
\end{eqnarray}
\subsection{Feasibility of the cloud control platform: Experiment of CCS based on PID controller}

To verify the feasibility of the platform, an experiment with PID controller is designed (see Algorithm \ref{alg:1}). If this platform can support PID controller, the data-driven method can also be applied to this platform.

\begin{algorithm}[!h]
    \renewcommand{\algorithmicrequire}{\textbf{Input:}}
	\renewcommand{\algorithmicensure}{\textbf{Output:}}
    \caption{\textbf{CCS Based on PID Controller}}
    \label{alg:1}
    \textbf{Input:} PID parameters $K_p$, $K_i$, $K_d$ and reference value $r_f$. \\
    \textbf{Output:} Control variable $u(k)$ by the cloud control service provider.
    \begin{algorithmic}[1]
        \STATE \textbf{\textit{Edge Client:}} At sampling time $k$, obtain the detected the ball's position $y(k)$. Then packet the data and send it to the cloud server.

        \STATE \textbf{\textit{Cloud Server:}} Receive the packet from the edge and compute control variable $u(k+1)$. Then packet the data and send it to the edge client.
    \end{algorithmic}
    \label{algorithm}
\end{algorithm}
\begin{figure*}[t]
\subfigure[]{
\begin{minipage}[b]{0.33 \textwidth}
\includegraphics[width=\textwidth]{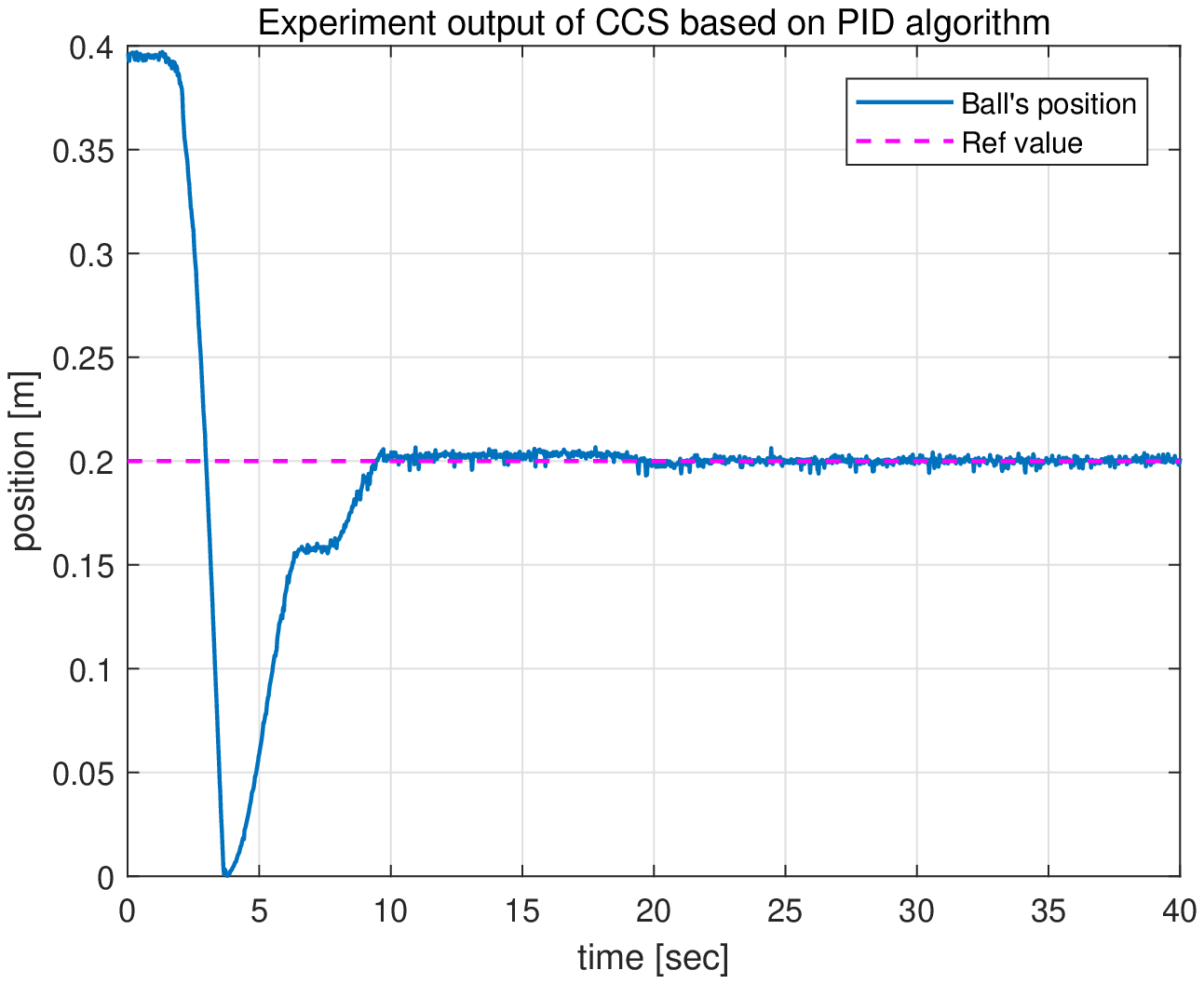}
\label{Experiment output of CCS based on PID algorithm}
\end{minipage}
}
\hspace{-0.2in}
\subfigure[]{
\begin{minipage}[b]{0.33 \textwidth}
\includegraphics[width=\textwidth]{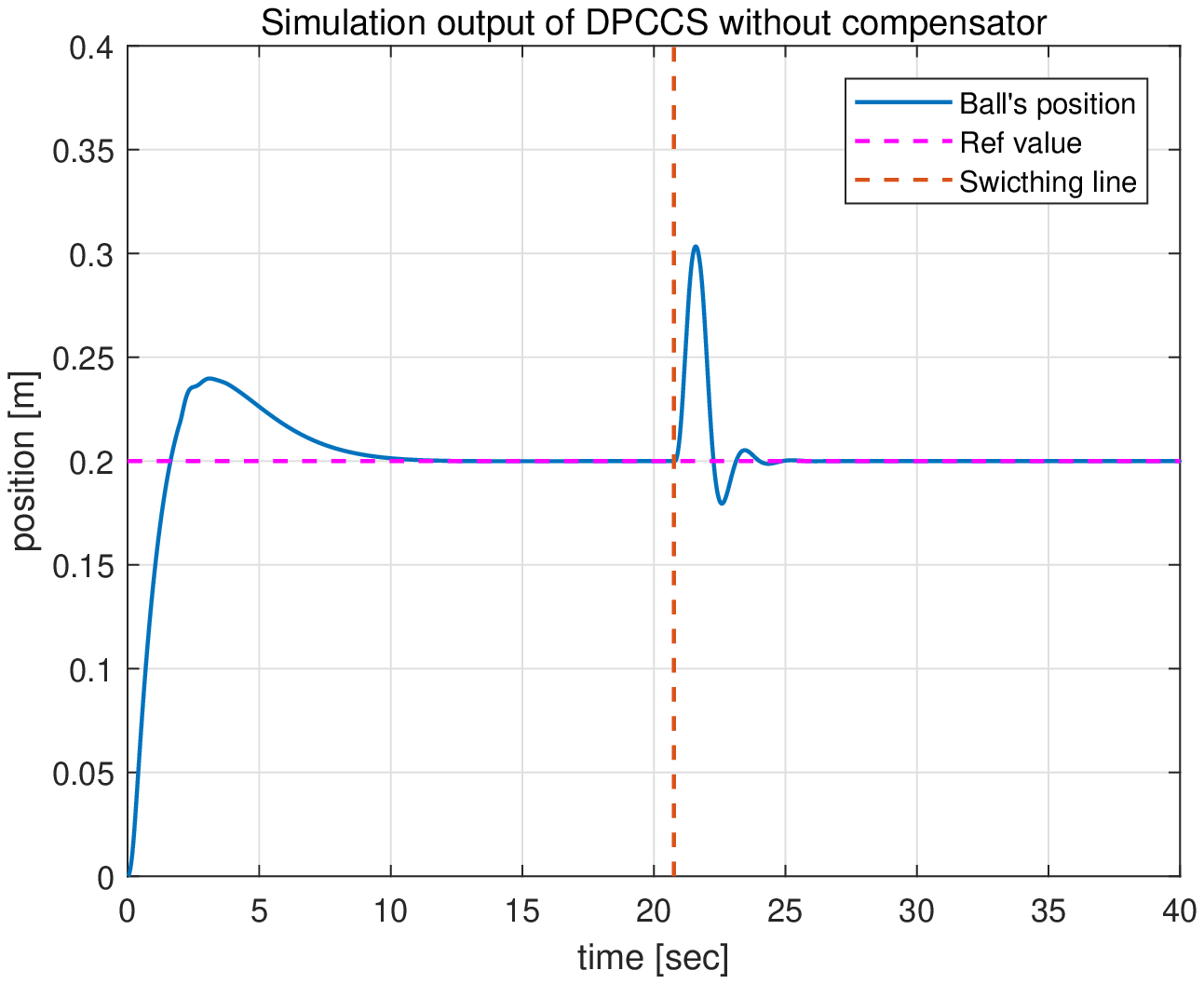}
\label{Simulation output of DPCCS without compensator}
\end{minipage}
}
\hspace{-0.2in}
\subfigure[]{
\begin{minipage}[b]{0.33 \textwidth}
\includegraphics[width=\textwidth]{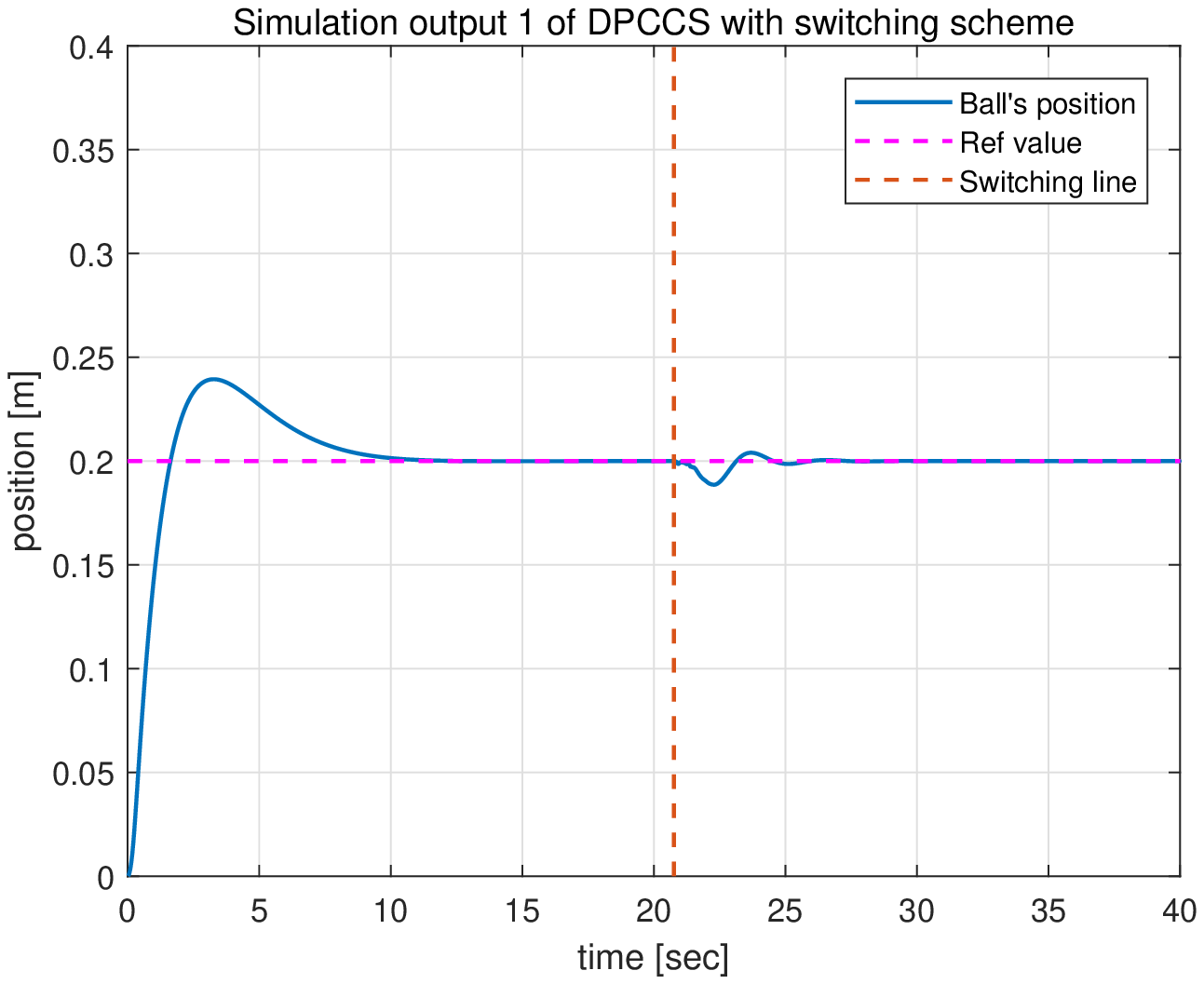}
\label{Simulation output 1 of DPCCS with switching scheme}
\end{minipage}
}

\hspace{-40.0pt}
\par \vspace{-10.pt}
\hspace{-36.0pt}

\subfigure[]{
\begin{minipage}[b]{0.33 \textwidth}
\includegraphics[width=\textwidth]{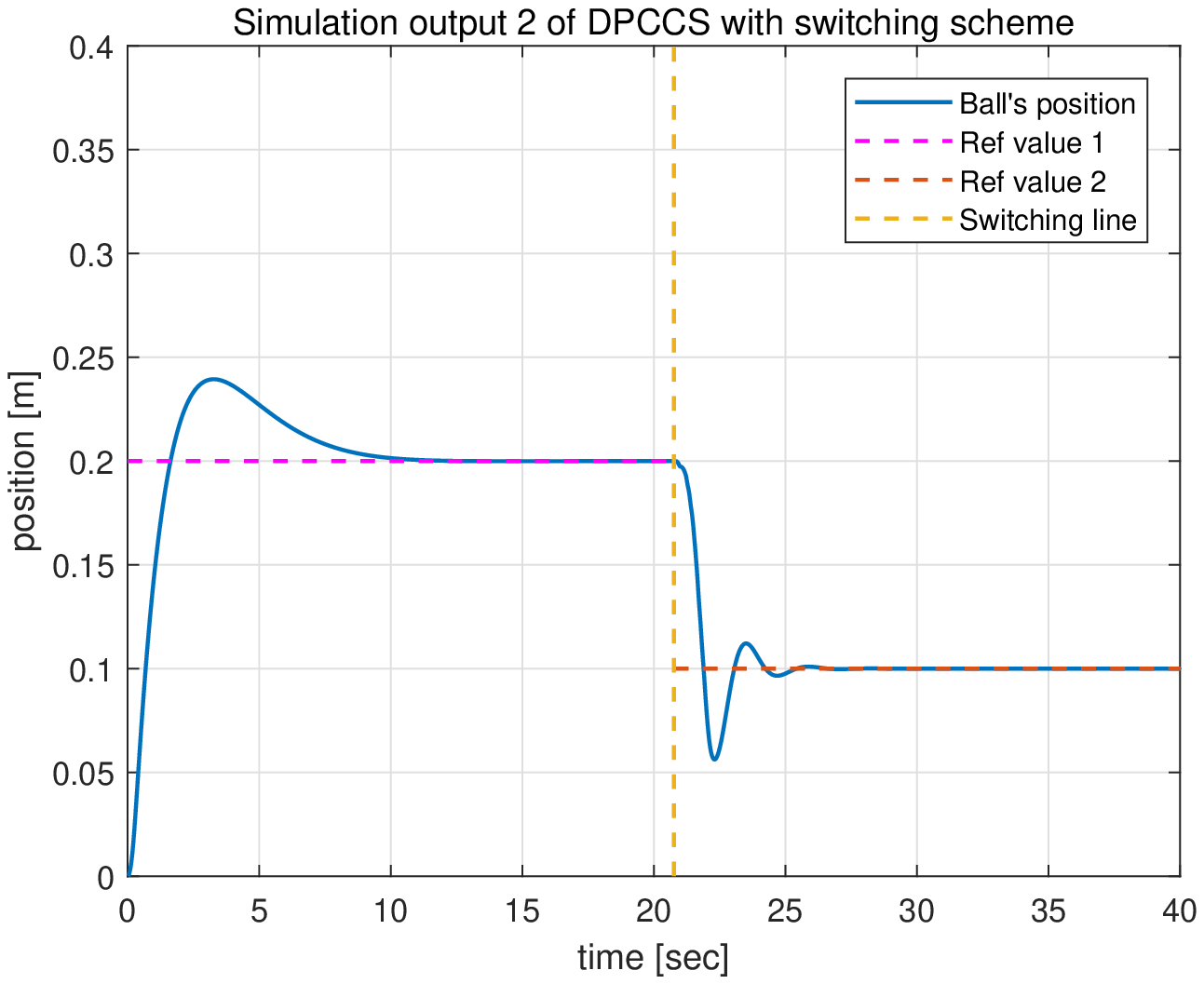}
\label{Simulation output 2 of DPCCS with switching scheme}
\end{minipage}
}
\hspace{-0.2in}
\subfigure[]{
\begin{minipage}[b]{0.33 \textwidth}
\includegraphics[width=\textwidth]{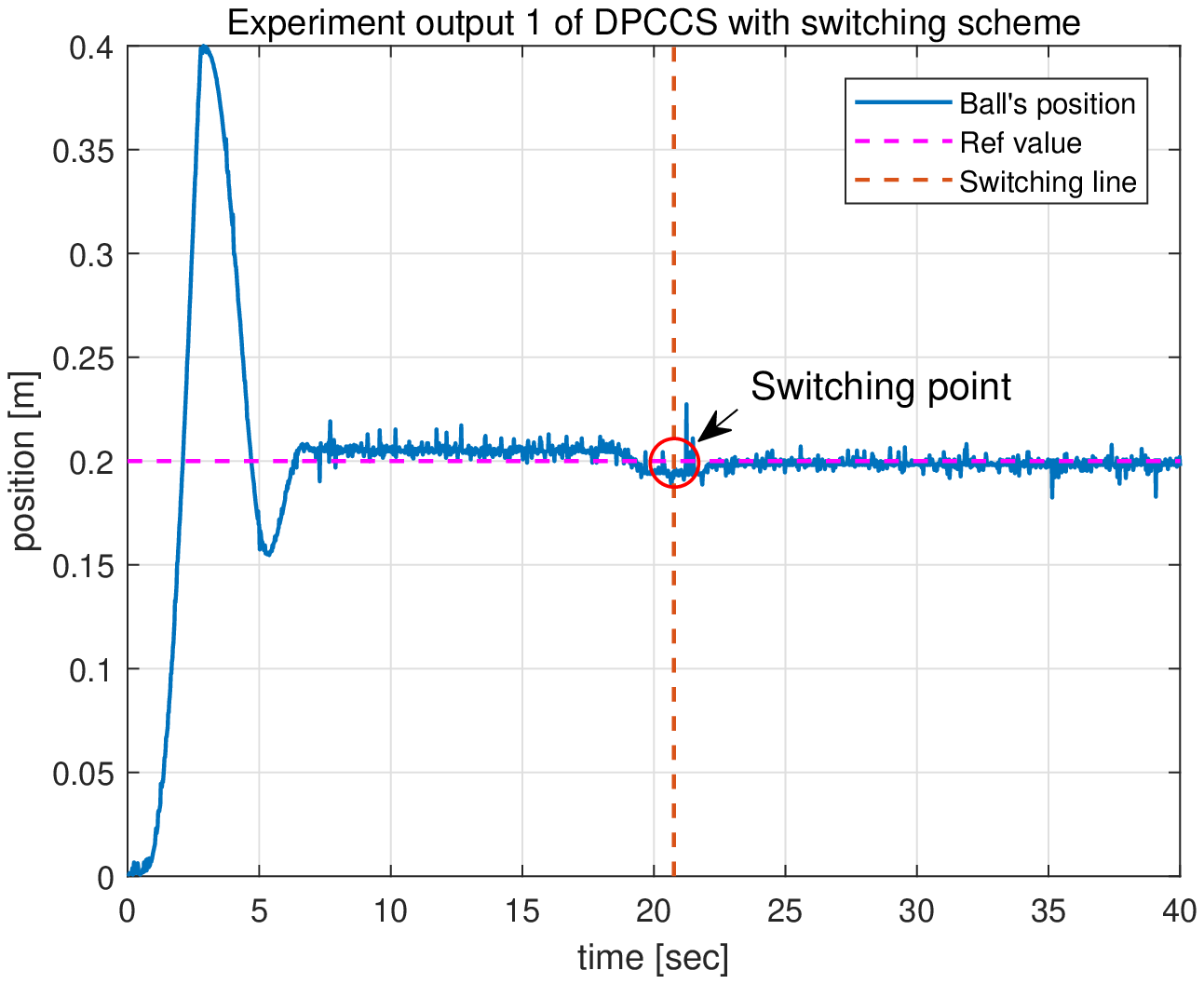}
\label{Experiment output 1 of DPCCS with switching scheme}
\end{minipage}
}
\hspace{-0.2in}
\subfigure[]{
\begin{minipage}[b]{0.33 \textwidth}
\includegraphics[width=\textwidth]{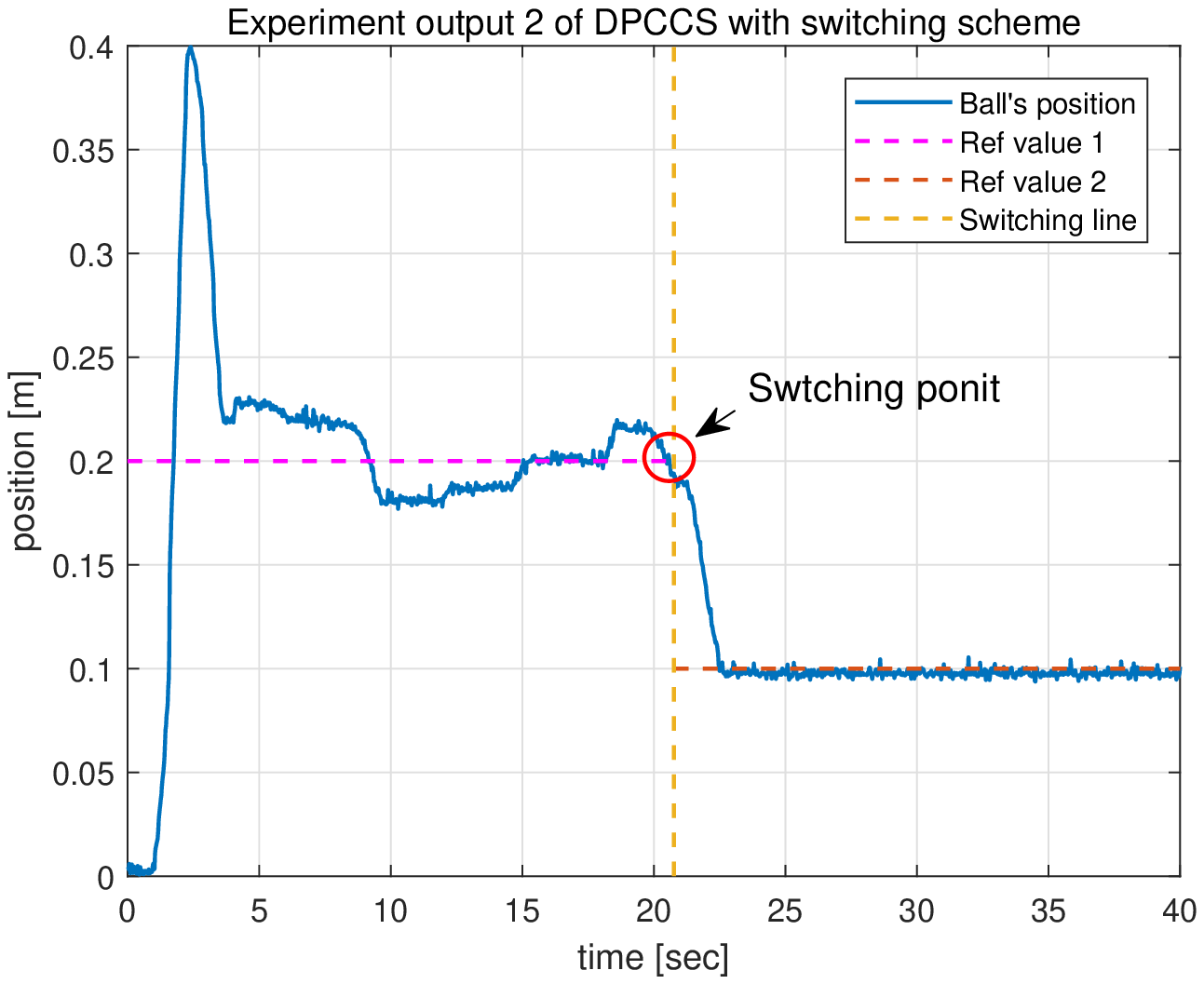}
\label{Experiment output 2 of DPCCS with switching scheme}
\end{minipage}
}
\caption{Simulations and Experiments of the Data-driven Predictive Cloud Control System (DPCCS)}
\end{figure*}
Placing the ball-beam system and edge client in Beijing and a PID cloud controller in Tencent Cloud, a CCS based on PID controller is established. In this experiment, the PID parameters are set as $(9.0, 3.0, 7.5)$ and the reference is set as 0.2 $m$. In Fig. \ref{Experiment output of CCS based on PID algorithm}, this platform shows desirable performance despite of a little disturbances occurring. The results demonstrate that the platform is feasible for CCS.

\subsection{Cloud communication compensator verification}

This part aims to test the validity of the cloud-edge compensator by performing two simulations. One is implemented with the compensator and another without it. The data-driven predictive cloud control algorithm is applied to this platform. The sampling period is set as 0.02 s. In the two simulations, we deploy this ball-beam system attached with a Matlab module and edge client in Beijing and the data-driven predictive controller in Tercent Cloud.

We first evaluate the method without the compensator and the result is depicted in Fig. \ref{Simulation output of DPCCS without compensator}. When the controller is switched from PID to data-driven method, some sharp shakes occur in the output curve for a while. Finally the output becomes stable at the reference value. The case shows the time delays bring negative impacts on performance intuitively, but this system can adjust the output stable at the reference value again by itself. This is because the features of time delays are considered into the implicit model in the data-driven control process. In the second case, the compensator is applied into the edge client. In Fig. \ref{Simulation output 1 of DPCCS with switching scheme}, the output curve is smoother, which illustrates the system employing the compensator works better with the same system parameters.

\vspace{-0.08in}
\subsection{Data-driven predictive cloud control system verification}

In Algorithm \ref{Data-driven Predictive cloud control system Algorithm}, preparing initial information is required to compute the original predictive control sequence. Therefore, this verification is designed as a switching control strategy. Especially, the first period is driven by PID controller and the second stage is controlled by the data-driven method. When the amount of initial data reaches a certain number, the data-driven method starts up. Two groups of simulations and experiments are performed to verify the effectiveness of the proposed system.

In the simulations, both the edge client and the controlled model are located at laboratory in Beijing and the controller in Tencent Cloud. As shown in Fig. \ref{Simulation output 1 of DPCCS with switching scheme}, PID controller provides the initial information. When both the numbers of input and output reach $2N+j$, some small shakes appear for nearly 3 seconds with the switching action. Soon after the shakes, this system is controlled smoothly until being stable. When the reference is changed at the switching moment, this system is also controlled well as shown in Fig. \ref{Simulation output 2 of DPCCS with switching scheme}.

In the experiments, the conditions are the same as those in the above simulations except that the controlled plant is a real ball-beam system. From Figs. \ref{Experiment output 1 of DPCCS with switching scheme} and \ref{Experiment output 2 of DPCCS with switching scheme}, we can obtain similar results with the simulations while the small burrs led by noise appear in experiments. According to the above results, the proposed data-driven predictive cloud control system demonstrates desirable effectiveness whether in simulations or experiments.

\begin{remark}
The data-driven control method in this work is obtained based on linear systems. This method implicitly updates the model and treats a control problem as a linear problem in each narrow time window. Therefore, this method receives desired performance on the nonlinear ball-beam system.
\end{remark}

\section{Conclusion}\label{Conclusion}

In this paper, an original data-driven predictive cloud control system is proposed for complex model-free control problems. A series of techniques are designed to establish a practical cloud control platform not a numerical simulator. The effectiveness of the proposed system is verified based on this practical platform. In the future, the inside features of cloud computing such as computing power, distributed ability and the ability of dynamical resource configuration would be brought into the proposed system and general cloud control system for further improvement.


\bibliographystyle{ieeetr}
\bibliography{IEEEabrv,mybib}

\end{document}